\documentclass[a4,12pt]{article}
\usepackage{graphicx}
\usepackage{amssymb}
\usepackage{slashed}
\usepackage{feynmf}
\usepackage{braket}
\usepackage{empheq}
\usepackage{dcolumn}
\usepackage{color}
\usepackage{comment}
\usepackage[pagebackref=false, colorlinks=true]{hyperref}
\hypersetup{linkcolor=blue, citecolor=blue, urlcolor=blue}
\usepackage{geometry}
\usepackage{authblk}
\usepackage[font=small]{caption}
\usepackage[font=small]{subcaption}
\newcommand{\be}{\begin{eqnarray}}
\newcommand{\ee}{\end{eqnarray}}
\newcommand{\nn}{\nonumber}
\newcommand{\f}{\frac}
\newcommand{\p}{\partial}

\usepackage{tikz}
\usepackage{tikz-feynman}
\usetikzlibrary{arrows.meta,arrows}

\title{Perturbative corrections to soft photon theorems for massless scalar QED in de Sitter spacetime}

\author[a]{Pratik Chattopadhyay\thanks{pratikpc@gmail.com}}
\affil[a]{School of Physics, The University of Electronic Science and Technology of China, No.2006, Xiyuan Avenue, West Hi-Tech Zone, Chengdu, Sichuan, P.R.China, Post Code: 611731}

\begin{document}
\maketitle

\begin{abstract}
The perturbative corrections to soft photon theorems with massive scalars in de Sitter spacetime were computed in \cite{Sayali_Diksha}. However, the massless limit of the scalar modes is ill-defined in their work. It therefore is ambiguous to take the massless limit of the soft factors. In this paper, we derive the massless scalar modes in $d$-dimensional de Sitter spacetime and use it to compute the perturbative corrections to the leading and sub-leading soft photon theorems. Our framework corresponds to tree level scattering of massless scalars followed by an emission of a soft photon in a compact region of the static patch of de Sitter. We show that our results are consistent with \cite{Sayali_Diksha} in the massless limit and we comment on the universality of our results.  
\end{abstract}

\section{Introduction}
Soft factorization properties of field theory amplitudes is known since a long time, both in electrodynamics and in gravity \cite{Bloch_1937, Gell-Mann_Goldberger, Low_1954, Low_1958, Weinberg_1965, Gross_Jackiw, Jackiw_1968, White_2011}. These properties capture the infrared behavior of amplitudes consisting of photons or gravitons, interacting with scalars or fermions. In this setup, one considers emission of an external photon or graviton from an external or internal leg of a scalar/fermion. In the soft limit of the emitted photon or graviton, the full amplitude factorizes into two parts, one that encodes the momentum and polarization of the soft particle and another that embeds the rest of the amplitude which, in general terminology, is referred to as the 'hard' part. This gives rise to leading, sub-leading and higher order soft factors. During the past decade, there has been a flurry of activities in this area, which started due to the seminal work of Strominger and collaborators \cite{Temple_He} and then by many others \cite{Laddha_2015, Daniel_Kapec, Lysov_Pasterski_Strominger, Campiglia_Laddha_2016}, who established the relation between the so called soft theorems and asymptotic symmetries. It was established that the soft theorems are nothing but the Ward identities of the asymptotic symmetries in gauge theory and gravity. There is another interesting phenomenon tied to these two aspects and is called the memory effect. It was shown that the memory effect is intimately connected with the soft theorems and asymptotic symmetries, thereby forming the famous infrared triangle \cite{Strominger_Zhiboedov, Pasterski_Strominger_Zhiboedov, Strominger}. Memory effects have also been studied in the context of near-horizon asymptotic symmetries, see \cite{Donnay_2016, Donnay_PRL_2016, Bhattacharjee_2020}.
\\~\\
Besides these activities, there has been growing interest to understand the soft theorems in curved spacetimes and their connections with the asymptotic symmetries in such spacetimes. For instance, significant progress have been made on the soft theorems in the context of anti-de Sitter space \cite{Fernandes_review_2023, Banerjee_2021, Banerjee_Bhattacharjee_2021, Banerjee_2023}. While this is already interesting, one has to ponder as to what is the fate of soft theorems in the context of de Sitter spacetime. De Sitter spacetime governs the inflationary scenario in the early universe and the late time accelerated universe approximately. Thus, studying soft theorems and asymptotic symmetries in this context is of paramount importance. In fact, there has been recent studies which show that in the small cosmological constant limit, the leading soft factor undergoes perturbative corrections due to the background de Sitter geometry \cite{Sayali_Diksha, SCB}. In addition, the soft theorems in the full static patch of de Sitter space have been derived from the Ward identities of the near-horizon symmetries in \cite{Mao_Zhou, Mao_Zhang}.
\\~\\
In \cite{Sayali_Diksha}, the analysis was done using massive scalars coupled to Maxwell field in generic dimensions. The massless limit in this case is not well defined as can be seen from their scalar mode solutions. Also, in their analysis, they have not considered the perturbative corrections to the sub-leading soft photon theorem. In this paper, we consider massless scalars coupled to photons in arbitrary dimensions and compute the perturbative corrections to the leading and sub-leading soft photon factors. The massless scalar field solution and the modes were derived in \cite{PD}, but in four dimensions. We generalize this to arbitrary dimensions and derive the propagator and the modes.  To make things consistent, we consider a double scaling limit, where the energy of the photon is taken to be very low and the de Sitter length scale is taken to be very high, in a manner such that their product is much much greater than unity.  
\\~\\
The paper is organized as follows: In section (\ref{setup}) we highlight the setup of our scattering problem and discuss the key assumptions made. In (\ref{gf}) we briefly study the U(1) gauge field in generic dimensions, which has already been discussed at length in \cite{Sayali_Diksha}. In (\ref{scalars_sec}), we study the massless scalar field in the de Sitter background in arbitrary dimensions. We find an orthogonal set of mode solution to the massless scalar field equation in the limit of large curvature length. We then use these set of modes to define the scalar field propagator and verify that it is the Green's function of the massless scalar field equation in section (\ref{prop_sec}). We then study the massless scalar field minimally coupled to U(1) gauge field in de Sitter background, and define the $\mathcal{S}$-matrix for the scattering  process. Evaluating the $\mathcal{S}$-matrix, we determine the perturbative corrections to the soft photon theorem in section (\ref{perturbative_correction_sec}) and summarize the results in section (\ref{results_sec}).
\\~\\

\section{Setup}
\label{setup}
\subsection{De Sitter spacetime}
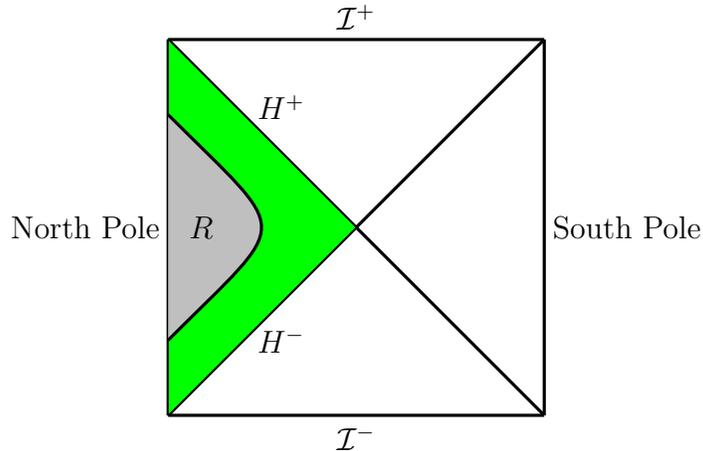
\begin{figure}
\centering
\begin{tikzpicture}
\draw[black, very thick] (-2.5,2.5) to node[shift={(0,0.3)}, sloped] {$\mathcal{I}^+$} (2.5,2.5);
\draw[black, very thick] (-2.5,-2.5) to node[shift={(0,-0.3)}, sloped] {$\mathcal{I}^-$} (2.5,-2.5);
\draw[black, very thick] (-2.5,2.5) to node[shift={(-1.1,0)}] {North Pole} (-2.5,-2.5);
\draw[black, very thick] (2.5,2.5) to node[shift={(1.1,0)}] {South Pole} (2.5,-2.5);
\draw[black, very thick] (-2.5,2.5) to node[shift={(-1,1.6)}] {$H^+$} (2.5,-2.5);
\draw[black, very thick] (2.5,2.5) to node[shift={(-1,-1.5)}] {$H^-$} (-2.5,-2.5);
\draw[fill=green] (-2.5,2.5) -- (0,0) -- (-2.5,-2.5) -- cycle;
\draw[black, very thick,fill=lightgray] (-2.5,1.5) to [out=-45, in=45,looseness=2] node[shift={(-0.8,0)}] {$R$} (-2.5,-1.5);
\end{tikzpicture}
\caption{Penrose diagram of de Sitter space.}
\label{dS_Penrose}
\end{figure}

We study the scattering of massless scalars followed by an emission of a photon in the static patch of the de Sitter space. We confine this scattering process to the small compact region $R$ inside the static patch as shown in Fig. (\ref{dS_Penrose}). The de Sitter metric can be put in the conformally flat form in the stereographic coordinates $x^\mu$ as is shown in \cite{SCB, Sayali_Diksha}, i.e,
\be 
g_{\mu\nu}=\Omega^2\eta_{\mu\nu}, ~~\Omega=\frac{1}{1+x^2/4l^2},
\ee
where $\eta_{\mu\nu}$ is the Minkowski metric. Inside the compact region, $x^\mu << l$, we can approximate the de Sitter metric up to $\mathcal{O}(l^{-2})$ as follows:
\be 
\label{metric}
g_{\mu\nu}\approx\Big(1-\frac{x^2}{2l^2}\Big)\eta_{\mu\nu}.
\ee
Thus, we effectively work in the large curvature length or small cosmological constant limit. We now discuss the key assumptions on our setup and various scales involved in the scattering process.
\subsection{Key assumptions in the setup}
\begin{itemize}
\item As is shown in Fig. (\ref{dS_Penrose}), we confine our scattering process in the region $R$ of the static patch in de Sitter spacetime. This region is much much smaller than the de-Sitter length scale $l$ and therefore we treat $l$ as a perturbation parameter in which the full scattering amplitude is expanded. Since the region $R$ is finite, particles can reach the boundary in a finite time $T$, which should be much larger than the interaction time scale. Thus, the particles are free from interactions at early and late times. To make this statement precise, one can define free particle states on early- and late-time Cauchy slices, which are the Hilbert spaces of the incoming and outgoing states respectively.
\item The $\mathcal{S}$-matrix for the scattering of massless scalars of momenta $p_1,...,p_n$ with an emission of a photon of momentum $k\ (=\omega\hat{k})$ decomposes in the soft limit $\omega\to 0$ as follows\cite{PD}:
\begin{equation} \label{Gamma_structure}
    \Gamma_{n+1}(\{p_1,...,p_n\},\omega\hat{k}) = A(\{p_i\},\omega\hat{k})\  \Gamma_{n}(\{p_1,...,p_n\}),
\end{equation}
\begin{equation}
    A(\{p_i\},\omega\hat{k}) = A^{\text{(flat)}}(\{p_i\},\omega\hat{k}) + \dfrac{1}{l^2} A^{\text{(dS)}}(\{p_i\},\omega\hat{k}),
\end{equation}
where $\Gamma_n$ is the $\mathcal{S}$-matrix for the scattering of massless $n$-scalars without an emission of the photon, $A^{(\text{flat})}$ is the Weinberg soft factor, and $A^{(\text{dS})}$ is the de Sitter correction.
These perturbative corrections are only valid for $\omega l >>1$. Also, the soft limit in the flat space requires $\omega << E$, where $E$ is the energy of the hard particles, i.e., scalars, that participate in the scattering process. Thus, in the present scenario, the soft limit is valid in the regime of $\frac{1}{l}<<\omega << E$. Since $l$ is very large and $\omega\to 0$, it is convenient to introduce a constant parameter $\delta=\omega l$. In this new parameter, the total soft factor can be expanded as follows:
\begin{equation}
    A(\{p_i\},\omega\hat{k}) = A^{\text{(flat)}}(\{p_i\},\omega\hat{k}) + \dfrac{1}{\delta^2} \tilde{A}^{\text{(dS)}}(\{p_i\},\omega\hat{k}),
\end{equation}
where $\tilde{A}^{\text{(dS)}} = \omega^2 A^{\text{(dS)}}$.
\end{itemize}
\section{Gauge field in de Sitter spacetime}\label{gf}
After the setup of the scattering process, we come to describing the U(1) gauge field in this spacetime. This description is similar to \cite{Sayali_Diksha}, where most of the derivations have been shown explicitly. Note, in $d=4$ the equations of motion for pure U(1) theory is the same as in flat spacetime. This is because U(1) gauge theory enjoys Weyl invariance. However, we do the analysis here in $d$ dimensions and thus the equations of motion for the gauge field will differ than that in flat spacetime. Also, in the presence of a current, the equations of motion admits a correction term of the order $\mathcal{O}(\frac{1}{l^2})$. In stereographic coordinates, the equation of motion reads
\be
\left(1+\frac{x^2}{2\ell^2}\right)\left(\Box A_\mu - \p_\mu(\p^\nu A_\nu)\ \right)\ +\ \f{(4-d)}{2\ell^2}[\ x.\p A_\mu -\p_\mu(x^\nu A_\nu)+A_\mu]\ =\ j_\mu.
\ee

Let us fix the gauge by setting
\be \label{gauge1}\p^\nu A_\nu
+\f{(4-d)}{2\ell^2}\ (x^\nu A_\nu)=0. \ee
This particular gauge fixing simplifies the equation of motion. Using the gauge condition \eqref{gauge1} in the equation of motion, we obtain
\be\label{geom}
\Box A_\mu \ +\ \f{(4-d)}{2\ell^2}\ [x.\p A_\mu +A_\mu]\ =\ j_\mu \left(1-\frac{x^2}{2\ell^2}\right).
\ee
\subsection{Photon modes}
The equation of motion \eqref{geom} admits homogenous solutions (i.e. $j_\mu = 0$) of the form 
\be \label{hk} f^h_{k\mu}(x) =\ \frac{1}{\sqrt{2 E_k}} \varepsilon_\mu^{h}(k)\ \left(1+\f{(d-4)}{8}\frac{x^2}{\ell^2}\right)\ e^{ik.x},\ \ k^2=\f{(d-2)(d-4)}{4\ell^2}, \label{Amodes} \ee
where $E_k$ is the zeroth component of $k^\mu$. Note that in generic dimensions, $k^2$ is non-zero, but if $d=4$, it is zero as expected.
After fixing all redundant degrees of freedom of the gauge field, one is left with $d-2$ physical degrees of freedom. Here we have used $h$ to denote these physical helicity states with $h$ taking $1$ to $d-2$.
\\
We define the Maxwell inner product on the solution space as
\begin{align}
(g^{\mu\nu}\ f^h_{k\mu},f^{h'}_{k'\nu})=&-i\int d^{d-1}x\sqrt{-g}\ g^{\mu\nu}(x)\ e^{-\epsilon \f{|\vec{x}|}{R}}\ [\ f^{*h}_{k\mu}(t,\vec{x})\ \p^t f^{h'}_{k'\nu}(t,\vec{x}) -f^{h'}_{k'\mu}(t,\vec{x})\ \p^t f^{*h}_{k\nu}(t,\vec{x})\ ]\nn\\
=&\ {(2\pi)^{d-1}} \delta_{h,h'}\ \delta^3(\vec{k}-\vec{k'}).\label{inA}
\end{align}
where we have added the exponential damping factor to get rid of the boundary terms. Hence the gauge field can be expanded in terms of the above modes as follows
\be
A_\mu(x)=\sum_{h=1}^{d-2}\int \f{d^{d-1}k}{(2\pi)^{d-1}} \ [\  a^h_k\ f_{k\mu}^{h}(x)\ +\  a^{h\dagger}_k \  f_{k\mu}^{*h}(x)\ ].\nn
\ee
The implication of the gauge condition \eqref{gauge1} in momentum space is
\be ik^\nu\varepsilon^h_\nu\  a^h_k\ e^{ik.x}
+\f{(4-d)}{4\ell^2}\  \varepsilon^h_\mu x^\mu\ a^h_k \ e^{ik.x}=0.\label{gauge}
\ee

\subsection{LSZ for photons}\label{gaugeap}
The LSZ formula for the photons have been derived explicitly in \cite{Sayali_Diksha}. We quote here the main points and the result. We first expand the gauge field as
\be
A_\mu(x)=\sum_{h=1}^{d-2}\int \f{d^{d-1}k}{(2\pi)^{d-1}}  \ [\ a^h_k f_\mu^{h}\ +\  a^{h\dagger}_k f_\mu^{* h}\ ].\nn
\ee
The above expression can be inverted using the orthogonality of modes using \eqref{inA} to obtain
\begin{align}
a^h_p=(g^{\mu\nu}\ f^{h}_{p\mu},\ A^{h'}_{k\nu}).
\end{align}
With this in hand, the derivation of the LSZ is straighforward and the final result is given as
\begin{align}
a^h_p(T)-a^h_p(-T)&=-i\int d^d x \sqrt{-g(x)}\eta^{\mu \nu} f^{*h}_{p\nu} \left(\left(1+ \f{x^2}{\ell^2}\right)\Box \ +\ \f{(4-d)}{2\ell^2}\ [x.\p +1]\right)A_\mu (x). \nn\\ 
\end{align}
For notational brevity, we define
\begin{equation}\label{Dx}
\mathcal{D}_x = \left(1+ \f{x^2}{\ell^2}\right)\Box+\ \f{(4-d)}{2\ell^2}\ [x.\p +1] ,    
\end{equation}
so that we have
\begin{align}\label{lszA}
a^h_p(T)-a^h_p(-T)&=-i\int d^d x \sqrt{-g(x)}\eta^{\mu \nu} f^{*h}_{p\nu} (\mathcal{D}_x) A_\mu (x).
\end{align}
We find that $\mathcal{D}_x$ acts on Greens function as follows:
\begin{align}\label{G11}
\mathcal{D}_x G_{\mu,\nu} (x,x') &= i\eta_{\mu\nu}\delta^d(x-x')\left[1+\f{dx^2}{4\ell^2}\right] \nn\\
&=  i\eta_{\mu\nu}\f{\delta^d(x-x')}{\sqrt{-g(x)}}
\end{align}

\section{Massless Scalars in de Sitter Background}
\label{scalars_sec}
\subsection{Scalar modes}
The modes for the massive scalar field in de Sitter background in the small cosmological constant limit have already been derived in \cite{Sayali_Diksha}, which is not well-defined in the massless limit. In this section, we derive the mode solutions for the massless scalars. The scalar field equation of motion in de Sitter background is 
\begin{equation}
\label{scalareom}
    \nabla^2\phi = 0,
\end{equation}
which in the small cosmological constant limit reduces to
\be \label{EOM}
\left(1+\frac{x^2}{2l^2}\right)\Box\phi +\frac{2-d}{2l^2}x\cdot\partial\phi=0,
\ee
where $\Box$ is the flat space d'Alembertian operator. The scalar field mode expansion can be written as
\be
\phi(x)=\int \f{d^{d-1}p}{(2\pi)^{d-1}} \ [a_p\ g_p(x)\ +\ b^\dagger_p\ g^*_p (x)],
\ee
where $E_p$ is the energy of the scalar field. Let us now choose the following ansatz to find a mode solution:
\be \label{ansatz}
g_p(x) = \dfrac{e^{ip\cdot x}}{\sqrt{2E_p}} \left(1 + \dfrac{\mathcal{F}(x)}{l^2} \right),
\ee
where $\mathcal{F}(x)$ is an arbitrary function which we want to determine. Substituting (\ref{ansatz}) into (\ref{EOM}) as a solution of the scalar field $\phi$, we obtain:
$$\mathcal{F}(x)=\dfrac{(d-2)x^2}{8},\ \ p^2=\dfrac{d-2}{l^2}.$$
As in \cite{Sayali_Diksha, SCB}, we define the inner product on the solution space as
\begin{equation}
\label{scalar_inner_product}
    \left( g_p, g_q \right) = -i \int d^{d-1}x \sqrt{-g} \ \big(g^*_p(x) \nabla^t g_q(x) - g_q(x) \nabla^t g^*_p(x) \big).
\end{equation}
Substituting the modes (\ref{ansatz}) in the above expression, one can verify that
\begin{equation}
\label{in}
    \left( g_p, g_q \right) = (2\pi)^{d-1} \delta^{(d-1)}(p-q).
\end{equation} 

We can express the creation and annihilation operators appearing above in terms of field operators by using the inner product defined in \eqref{in}. This deduction has been done in \cite{Sayali_Diksha} but in their case, it seems they have considered a neutral scalar field as their mode solution is real. This is not legitimate and hence we take a complex mode solution. Now the important point we is that the LSZ and the S matrix does not change if some of the scalars are anti particles instead of particles. This is due to crossing symmetry. Therefore, where confusions do not arise, we consider all scalars to be particles (and not antiparticles or a mixture). We then return to their result and quote it here. For the annihilation operator corresponding to the scalar field, we have
\begin{align}
a_p(T)-a_p(-T)&=\int^T_{-T} dt\ \p_t a_p\nn\\
&=-i\int^T_{-T} dt d^{d-1}x\ e^{-\epsilon \f{|\vec{x}|}{R}}\ \sqrt{-g}\ \nabla_\mu [\ g_p^*\nabla^\mu \phi -\phi\nabla^\mu g_p^*\ ]\ 
\end{align}
Using the fact $g_p^*$ satisfies the scalar equation of motion given in \eqref{scalareom} and taking $\epsilon \rightarrow 0$, we get
\begin{align}
a_p(T)-a_p(-T)&=-i\int^T_{-T} dt d^{d-1}x\ \sqrt{-g}\  g_p^*\ [\nabla^2]\ \phi .
\end{align}

\subsection{Scalar propagator}
\label{prop_sec}
The differential equation for the scalar field propagator is
\be
\nabla_x^2 D(x,y)=i \frac{\delta^{(d)}(x-y)}{\sqrt{-g}},
\ee
which can be expanded up to $\mathcal{O}(l^{-2})$ as
\be \label{propeom}
\left[\left(1+\frac{x^2}{2l^2}\right)\Box_x+\frac{2-d}{2l^2}x\cdot\p_x\right]D(x,y) = i \frac{\delta^{(d)}(x-y)}{\sqrt{-g}}.
\ee
We can write the symmetric form of the Feynman propagator using the orthogonal set of modes $g_p$ as
\begin{align} \label{scalar_propagator}
    D(x,y)&=-i\int\dfrac{d^dp}{(2\pi)^d} \dfrac{(\sqrt{2E_p})g_p(x)(\sqrt{2E_p})g^*_p(y)}{p^2-\frac{(d-2)}{l^2}-i\epsilon} \nonumber \\
    &=-i\int\dfrac{d^dp}{(2\pi)^d} \dfrac{e^{ip\cdot(x-y)}}{p^2} \left(1 + \dfrac{(d-2)x^2}{8l^2} + \dfrac{(d-2)y^2}{8l^2} + \dfrac{(d-2)}{p^2l^2} \right).
\end{align}
We substituted the above form of the scalar propagator (\ref{scalar_propagator}) into (\ref{propeom}) to verify that it is indeed the Green's function of the massless scalar field equation.
\section{Perturbative $\mathcal{S}$-matrix and soft factor corrections}
\label{S-matrix_sec}
In this section, we define a perturbative $\mathcal{S}$-matrix in the de Sitter space and deduce curved spacetime generalization of the LSZ formula. The LSZ formula in \cite{SCB} was obtained for massive scalars. Here, we focus on massless scalars. Thus, in a sense, the perturbative $\mathcal{S}$-matrix for the present case should be obtained in a suitable limit of the scalar mass, $m\rightarrow 0$. Another point to mention is that to formulate a convenient $\mathcal{S}$-matrix, we need to prescribe asymptotic states at the boundaries of spacetime. However, in de Sitter space, there is no notion of a global observable boundary and the only observable region is the static patch. This patch is bounded by the cosmological horizons. In this situation, we defer from defining proper asymptotic states at null infinity but instead treat the cosmological horizons as physical boundaries of this spacetime \footnote{In \cite{Teitelboim_2001, Teitelboim_2003}, the cosmological horizons are considered as physical boundaries to study the thermodynamics of a rotating black hole.} to define the so called asymptotic states. In our analysis, since the scattering processes are confined to the compact region inside the static patch, the asymptotic states are defined at the early- and late-time Cauchy surfaces which act as Hilbert spaces of the incoming and outgoing particles. The particles can reach the Cauchy surface in a finite time $T$, which we assume to be much larger than the interaction time scale. All lengths in our calculations are very small compared to the de Sitter radius $l$, i.e., $|x|<<l$.

The $\mathcal{S}$-matrix describing the scattering of incoming particles with momenta $p_i$ to the outgoing particles with momenta $p_j$ followed by the emission of photons with momenta $p_k$ is defined as 
\begin{equation}
\label{Gamma_general}
    \Gamma (\{p_j,p_k\},\{p_i\}) = \lim_{t\to T} \left\langle \mathcal{T} \prod_{j,k\in \text{out}} \sqrt{2E_{p_j}} a_{p_j}(t)\ \sqrt{2E_{p_k}} a_{p_k}(t) \prod_{i\in \text{in}} \sqrt{2E_{p_i}} a^{\dagger}_{p_i}(-t) \right\rangle,
\end{equation}
where $\mathcal{T}$ represents time-ordering.
To deduce the LSZ formula, we rewrite the above expression as follows:
\begin{multline}
\label{Gamma_2}
    \Gamma (\{p_j,p_k\},\{p_i\}) = \lim_{t\to T} \bigg\langle \mathcal{T} \prod_{j,k\in \text{out}} \sqrt{2E_{p_j}} \big(a_{p_j}(t) - a_{p_j}(-t)\big) \sqrt{2E_{p_k}}\big(a_{p_k}(t)-a_{p_k}(-t)\big) \\
    \cdot \prod_{i \in \text{in}} \sqrt{2E_{p_i}} \big(a^{\dagger}_{p_i}(-t)-a^{\dagger}_{p_i}(t)\big) \bigg\rangle,
\end{multline}
where all the annihilation (creation) operators go to the right (left) side in the time order product and annihilate the ``in" (``out") -state, so that the above expression reduces to Eq. (\ref{Gamma_general}). By inverting the inner product expression for the scalar field modes, one can deduce \cite{Sayali_Diksha}
\begin{equation}
    a_p(T)-a_p(-T) = -i\int_{-T}^{T} d^dx \sqrt{-g} g_p^*(x)\nabla^2\phi(x).
    \label{s_inner_invert}
\end{equation}
Substituting Eqs. (\ref{s_inner_invert}) and (\ref{lszA}) into Eq. (\ref{Gamma_2}) yields
\begin{align}
\Gamma(\lbrace p_i,p_k\}, \{p_j \rbrace)&=\int \prod_{i,k\in \text{in}}\prod_{j\in \text{out}}\ [d^dx_i] [d^dy_j] [d^dz_k] \  \eta^{\sigma\mu} g_{p_j}(y_j)g^*_{p_i}(x_i)f^{*h}_{p_k \sigma}(z_k)\ \nn\\
& (-i)[\nabla_i^2]\  (-i)[\nabla_j^2]\  (-i)\mathcal{D}_{z_k} \langle \mathcal{T}\ A_{\mu}(z_k)\phi({x_i})\ldots \phi({y_j})\rangle .\label{LSZ}
\end{align} 
where the measure factors are given by:
\be
[d^dx_i] [d^dy_j] [d^dz_k] 
 = d^dx_i\ d^dy_j\ d^dz_k \ \sqrt{2E_{p_i} (2E_{p_j})\ (2E_{p_k})} \ \sqrt{-g(x_i) \ (-g(y_j)) \ (-g(z_k))}. \nn 
\ee
\subsection{Soft photon scattering}\label{corrections}
In the previous sections, we found $O(\f{1}{\ell^2})$ corrections to the scalar modes and propagators of scalar fields and of the gauge fields. Let us now compute the perturbative corrections to the flat space soft photon theorems in generic dimensions. The action for a complex scalar field minimally coupled to the U(1) gauge field is given by
\begin{equation}
\mathcal{S} = -\int d^{d}x \sqrt{-g}\ \Big[\frac{1}{4}g^{\mu\rho}g^{\nu\sigma}F_{\mu\nu}F_{\rho\sigma} + g^{\mu\nu}\left(D_{\mu}\phi\right)^{*}\left(D_{\nu}\phi\right) +\ V[|\phi|^2]\  \Big],
\end{equation}
where
$$D_\mu\phi=\p_\mu\phi-ieA_\mu\phi$$
and $V[|\phi|^2]$ is a gauge invariant scalar potential. As is well known, the leading order (in $(\f{1}{k})$) soft photon theorem in flat spacetime establishes a relation between an amplitude with $n$ hard particles and one soft photon to the amplitude without the soft photon. It takes the following form
\be
\Gamma_{n + 1}(\{p_i\}, k) = A(\{p_i\} , k) \Gamma_{n}(\{p_i\}),
\ee
where $p_i$ are the momenta of hard particles, $k$ is the soft momentum and $A(\{p_i\} , k)$ is the soft factor.

\begin{figure}[h]
\centering
\begin{tikzpicture}[line width=1.5pt, scale=1.5]

\begin{scope}[xshift=0cm]
\begin{feynman}
\begin{scope}[shift={(6,0)}]
\draw[scalar] (.7,0)--(2,0);
\draw[scalar][rotate=30] (.7,0)--(2,0);
\draw[scalar][rotate=180] (.7,0)--(1.5,0);
\draw[scalar][rotate=180] (2.5,0)--(1.5,0);
\draw[scalar][rotate=150] (.7,0)--(2,0);

\draw[photon][rotate=180] (1.5,0)--(1.5,1.3);

\node at (-0.8,.7) {$\bullet$};
\node at (-0.2,.95) {$\bullet$};
\node at (0.5,.9) {$\bullet$};
\node at (-1.5,.2) {$z'$};
\node at (-.8,-.2) {$z$};
\node at (-1.5,-1.5) {$(\varepsilon_{\mu }, k)$};
\node at (-2.5,.2) {$p_i$};

\begin{scope}[shift={(0,0)}, scale=2]
    \draw [ultra thick] (0,0) circle (.35);
    \clip (0,0) circle (.3cm);
    \foreach \x in {-.9,-.8,...,.3}
        \draw[line width=1 pt] (\x,-.3) -- (\x+.6,.3);
\end{scope}

\end{scope}
\end{feynman}
\end{scope}

\begin{scope}[xshift=5cm]
\begin{feynman}
\begin{scope}[shift={(6,0)}]
\draw[scalar] (.7,0)--(2,0);
\draw[scalar][rotate=30] (.7,0)--(2,0);
\draw[scalar][rotate=180] (.7,0)--(1.5,0);
\draw[scalar][rotate=180] (2.5,0)--(1.5,0);
\draw[scalar][rotate=150] (.7,0)--(2,0);
\coordinate (blob) at (.7,0);

\coordinate (photonstart) at ($(blob) + (0,-0.2)$);  

\draw[photon] (photonstart) -- ++({1.7*cos(-30)}, {1.7*sin(-30)}) 
    node[pos=1, above right] {$(\varepsilon_\mu, k)$};

\node at (-0.8,.7) {$\bullet$};
\node at (-0.2,.95) {$\bullet$};
\node at (0.5,.9) {$\bullet$};
\node at (-1.5,.2) {};
\node at (-.8,-.2) {};
\node at (2.0,0.2) {$p_i$};

\begin{scope}[shift={(0,0)}, scale=2]
    \draw [ultra thick] (0,0) circle (.35);
    \clip (0,0) circle (.3cm);
    \foreach \x in {-.9,-.8,...,.3}
        \draw[line width=1 pt] (\x,-.3) -- (\x+.6,.3);
\end{scope}

\end{scope}
\end{feynman}
\end{scope}

\end{tikzpicture}
\caption{Left: photon emitted from external scalar; Right: photon emitted from an arbitrary internal line.}
\label{soft_graviton_fig}
\end{figure}
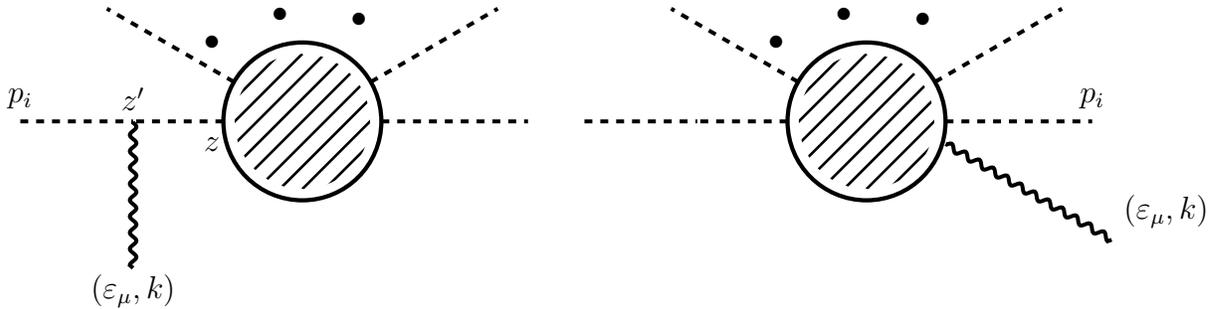

Due to the presence of de Sitter background, the flat space soft factor as well as the amplitude $\Gamma_{n}(\{p_i\})$ receives $O(\f{1}{\ell^2})$ corrections.
\\
Now, we are in a position to compute $\Gamma_{n + 1}(\{p_i\}, k)$ using the curved space LSZ formula discussed in \eqref{LSZ} of the previous section. However, this analysis is already done in \cite{Sayali_Diksha} and we quote only the final result. The main addition to what has already been done in 
\cite{Sayali_Diksha} is that we also have another diagram for the $\mathcal{S}$-matrix where the photon is attached to an internal line, see figure (\ref{soft_graviton_fig}). In the massless limit of the scalar, $m=0$, we have
\begin{align}\label{softf}
\Gamma_{n + 1}(\{p_1...p_n\}, k)\ &=  -(-i)^{n+1} \int d^dz\ d^dz' \left(1 - \frac{dz^2}{4\ell^2}- \frac{(d-2)z'^2}{4\ell^2}\right)  \sum_{i = 1}^{n}e_i\ g^*_{p_i}(z') f^{*}_{k\mu}(z')\ \eta^{\mu\nu}\nn\\
 &   \left( 2\p'_\nu D(z',z) \   -\frac{z'_{\nu}}{\ell^2}D(z',z)\right) \ \int \prod_{\substack{j = 1\\ j \neq i}}^{n-1}d^d z_j \sqrt{-g(z_j)}\ g^*_{p_j}(z_j))
 (\nabla_{z_j}^2)G[\{\phi(z_j)\}]\ \nn\\+ & M_{n+1}(\{p_1,...,p_n\},k),
 \end{align}
where $G[\phi(z)...\phi(z_j)]$ is the $n$-point correlation function which receives contributions from any arbitrary interaction vertices, $M_{n+1}$ is the $\mathcal{S}$-matrix for the second diagram in Fig. (\ref{soft_graviton_fig}) where the photon is not attached to an external line but to an internal line. The summation has been  taken over all the scalars since any of them can emit the photon.
This is the main result from which we compute the leading, subleading, and sub-subleading corrections to the flat space soft factor in the following sections. 

\subsection{Perturbative corrections to soft photon theorem}
\label{perturbative_correction_sec}
We now substitute the scalar modes (\ref{ansatz}), scalar propagator (\ref{scalar_propagator}), and photon modes (\ref{Amodes}) into Eq. (\ref{softf}) and obtain the following
\begin{align}
&\Gamma_{n+1}(\{p_1,...,p_n\},k) =\sum_{i=1}^{n}e_i
\int d^dz d^dz' \int\dfrac{d^dp}{(2\pi)^d} \bigg(\dfrac{p^\mu  \varepsilon_{\mu}}{p^2} - \dfrac{z'^2(p^\mu \varepsilon_{\mu})}{2l^2p^2}
- \dfrac{dz^2(p^\mu \varepsilon_{\mu})}{2l^2p^2} \nonumber \\
&+ \dfrac{2(d-2)p^\mu\varepsilon_{\mu}}{l^2p^4} + \dfrac{-i(d-2)z'^\mu \varepsilon_{\mu}}{4l^2p^2}\bigg) 
e^{i(p-p_i-k)\cdot z'} e^{-ip\cdot z} 
\int\prod_{j=1; j\ne i}^{n-1} d^dz_j \sqrt{-g(z_j)} g^*_{p_j}(z_j)(-i)(\nabla^{2}_{z_j}) \big\langle G[\{\phi(z_j) \}] \big\rangle \nonumber \\
&+ M_{n+1}(\{p_1,...,p_n\},k).
\end{align}
We now simplify the expression above. To do so, we replace $z'$ by $-i\partial_p$ and then perform integration by parts, thereby bringing the exponential factor $e^{ip\cdot z'}$ outside the derivative operator $\partial_p$. Collecting the exponential factors together, we then perform the $z'$ integral, yielding the delta function $\delta^{(4)}(p-p_i-k)$. Finally, we do the $p$-integral using the delta function support and use the soft limit in $k$. This gives us 
\begin{align}
&\Gamma_{n+1}(\{p_1,...,p_n\},k) = \sum_{i=1}^{n}e_i \int d^dz \bigg(\dfrac{p_i^\mu \varepsilon_{\mu}}{p_i\cdot k} - \dfrac{i(k\cdot z)(p_i^\mu\varepsilon_{\mu})}{p_i\cdot k} + \dfrac{(d-4)}{4l^2} \dfrac{p_i^\mu\varepsilon_{\mu}}{(p_i\cdot k)^2} \nonumber \\
&+\dfrac{i(d-4)}{4l^2} \dfrac{(p_i\cdot z)(p_i^\mu \varepsilon_{\mu})}{(p_i\cdot k)^2}
\bigg) e^{-ip_i\cdot z} \int\prod_{j=1; j\ne i}^{n-1} d^4z_j \sqrt{-g(z_j)} g^*_{p_j}(z_j)(-i)(\nabla^{2}_{z_j}) \big\langle G[\{\phi(z_j) \}] \big\rangle \nonumber \\
& + M_{n+1}(\{p_1,...,p_n\},k).
\end{align}
We now replace $z$ by $i\partial_{p_i}$, and put the above expression in the final form as
\begin{align} \label{Gamma=soft+N}
\Gamma_{n+1}(\{p_1,...,p_n\},k) &= \sum_{i=1}^{n}e_i\bigg[\dfrac{p_i^\mu\varepsilon_{\mu}}{p_i\cdot k} + \dfrac{p_i^\mu\varepsilon_{\mu}}{p_i\cdot k} k\cdot\partial_{p_i} + \dfrac{(d-4)}{4l^2} \dfrac{p_i^\mu\varepsilon_{\mu}}{(p_i\cdot k)^2}\nonumber\\& - \dfrac{(d-4)}{4l^2} \dfrac{p_i^\mu\varepsilon_{\mu}}{(p_i\cdot k)^2} p_i\cdot\partial_{p_i}
\bigg] \Gamma_{n}(p_1,...,p_n) \nonumber \\
&+ M_{n+1}(\{p_1,...,p_n\},k).
\end{align}
In the above equation, $\Gamma_n$ is the $\mathcal{S}$-matrix for the scattering of massless $n$-scalars without an emission of the photon. The above expression should be invariant under the gauge transformation,
\begin{equation}
    \varepsilon_{\mu} \to \varepsilon_{\mu} + ik_\mu + \mathcal{O}(l^{-2}).
\end{equation}
If the $\mathcal{S}$-matrix for the scattering process where the photon is attached to an internal line is defined by $M_{n+1}(\{p_1,...,p_n\},k)$, then it can be decomposed in the soft expansion as follows:
\begin{multline}
    M_{n+1}(\{p_1,...,p_n\},k) = 
    M^{(0)}_{n+1}(\{p_1,...,p_n\},\omega \hat{k}) + \frac{1}{\ell} M^{(1)}_{n+1}(\{p_1,...,p_n\}, \omega\hat{k})\\
    + \frac{1}{\ell^2} M^{(2)}_{n+1}(\{p_1,...,p_n\},\omega \hat{k}).
\end{multline}
Since the soft photon is not attached to an external line, this diagram does not produce $\frac{1}{\omega}$-pole terms. Therefore, the above expression can be expanded at $\omega\to 0$ as follows:
\begin{multline}
    M_{n+1}(\{p_1,...,p_n\},k) = M^{(0)}_{n+1}(\{p_1,...,p_n\}, 0) + \frac{1}{\ell}M^{(1)}_{n+1}(\{p_1,...,p_n\}, 0)\\
    + \frac{1}{\ell^2}M^{(2)}_{n+1}(\{p_1,...,p_n\}, 0) + ...,
\end{multline}
where $M^{(0)}_{n+1}$ produce the flat space subleading soft factor, $M^{(1)}_{n+1}$ and $M^{(2)}_{n+1}$ are correction terms.
We now add the additional correction terms to Eq. (\ref{Gamma=soft+N}). The resultant expression is
\begin{align}
\label{soft+M}
\Gamma_{n+1}(\{p_1,...,p_n\},k)& = \sum_{i=1}^{n}e_i\bigg[\dfrac{p_i^\mu\varepsilon_{\mu}}{p_i\cdot k} + \dfrac{p_i^\mu\varepsilon_{\mu}}{p_i\cdot k} k\cdot\partial_{p_i} + \dfrac{(d-4)}{4l^2} \dfrac{p_i^\mu\varepsilon_{\mu}}{(p_i\cdot k)^2} \nonumber\\&- \dfrac{(d-4)}{4l^2} \dfrac{p_i^\mu\varepsilon_{\mu}}{(p_i\cdot k)^2} p_i\cdot\partial_{p_i}
\bigg] \Gamma_{n}(p_1,...,p_n)\nonumber\\
&+\varepsilon_{\mu} M^{(0)\mu}_{n+1}(\{p_1,...,p_n\}, 0) + \delta \varepsilon_{\mu} M^{(1)\mu}_{n+1}(\{p_1,...,p_n\}, 0),
\end{align}
where the second term arises from the soft expansion of the exponential factor $e^{-ik\cdot z'}$, and $M_{n+1}$ is decomposed as $\varepsilon_{\mu}M_{n+1}^{\mu}$.
Now we use gauge invariance to evaluate the additional correction term $M^{(0)\mu}_{n+1}$. At the leading order, the gauge transformation implies $\varepsilon_{\mu} \to \varepsilon_{\mu} + ik_\mu$. The photon polarization will therefore shift under this transformation because at this order, it doesn't receive $\mathcal{O}(\f{1}{\ell^2})$ corrections, which would otherwise render the gauge transformation to be complicated. Since the $\mathcal{S}$-matrix should remain invariant, we have, 
\begin{equation}
\label{gauge_invariance_S-matrix}
    \sum_{i=1}^n e_i\bigg(1 + k\cdot\partial_{p_i}\bigg) \Gamma^{(0)}_n
    +k_\mu M^{(0)\mu}_{n+1} = 0,
\end{equation}
where the first term vanishes due to charge conservation. Equating the remaining term at the leading order, we obtain
\begin{align}
    &M_{n+1}^{(0)\mu} = -\sum_{i=1}^{n} e_i\frac{\partial}{\partial_ {p_i}^\mu} \Gamma_n^{(0)}, \label{M0}
\end{align}
Substituting the above into (\ref{Gamma=soft+N}) and simplifying it, we obtain the final expression as follows:
\begin{align}
\Gamma_{n+1}(\{p_1,...,p_n\},k) &= \sum_{i=1}^{n}e_i\bigg[\dfrac{p_i^\mu\varepsilon_{\mu}}{p_i\cdot k} - i\dfrac{\varepsilon_{\mu}k_\nu J_i^{\mu\nu}}{p_i\cdot k} + \dfrac{(d-4)}{4l^2} \dfrac{p_i^\mu \varepsilon_{\mu}}{(p_i\cdot k)^2} \nonumber\\&- \dfrac{(d-4)}{4l^2} \dfrac{p_i^\mu \varepsilon_{\mu}}{(p_i\cdot k)^2} p_i\cdot\partial_{p_i}
\bigg] \Gamma_{n}(p_1,...,p_n) \nonumber \\
&+ \dfrac{1}{l^2} M^{(2)}_{n+1}(\{p_1,...,p_n\},k),
\end{align}
where $J^{\mu\nu}=i\big(p^\mu\partial^\nu_p - p^\mu\partial^\nu_p \big)$ is the angular momentum operator.
We now introduce the parameter $\delta=\omega l$ in the soft limit $\omega\to 0$. Therefore,
\begin{align}
\label{Gamma_in_delta_1}
\Gamma_{n+1}(\{p_1,...,p_n\},\omega \hat{k}) &= \sum_{i=1}^{n} e_i\bigg[\dfrac{p_i^\mu\varepsilon_{\mu}}{\omega p_i\cdot \hat{k}} - i\dfrac{\varepsilon_{\mu}\hat{k}_\nu J_i^{\mu\nu}}{p_i\cdot \hat{k}} + \dfrac{(d-4)}{4\delta^2} \dfrac{p_i^\mu \varepsilon_{\mu}}{(p_i\cdot \hat{k})^2} \nonumber\\&- \dfrac{(d-4)}{4\delta^2} \dfrac{p_i^\mu \varepsilon_{\mu}}{(p_i\cdot \hat{k})^2} p_i\cdot\partial_{p_i}
\bigg] \Gamma_{n}(p_1,...,p_n) \nonumber \\
&+ \dfrac{1}{\delta^2} M^{(2)}_{n+1}(\{p_1,...,p_n\},\hat{k}),
\end{align}
This is the full $\mathcal{S}$-matrix at this order, and it agrees with that of \cite{Sayali_Diksha} in the limit where the scalar mass is zero. The only addition here is the contribution from the photon attached to an internal line and this contribution is the second term of the above equation. In the above expression the momenta $p_i$ and $k$ are defined on the de Sitter background and they obey the dispersion relations $p_i^2=\f{(d-2)}{l^2}, k^2=\f{(d-2)(d-4)}{4l^2}$.
\section{Results}
\label{results_sec}
Let us now write down the results which we achieved. However, there is one subtlety. Our previous expression of the $S$-matrix was in terms of curved space momenta.
When we decompose both the scalar and gauge field momenta in de Sitter into flat spacetime, some hidden correction terms emerge. Let us then compute these correction terms which arise from the leading and sub-leading soft factors. First, using the dispersion relations, we can parametrize the de Sitter momenta as follows:
\begin{align}
    &k^\mu =  \omega \left(1, \left(1+\dfrac{\lambda}{2\delta^2} \right){\mathbf{\hat{k}}} \right), \\
    &p_i^\mu = E_i \left(1,\ \left(1+\dfrac{(d-2)\omega^2}{2\delta^2 E_i^2} \right) {\mathbf{\hat{p_i}}} \right),
\end{align}
where $\mathbf{\hat{k}}$ and $\mathbf{\hat{p_i}}$ are the unit $(d-1)$-vectors that parametrize points on the $(d-2)$-dimensional sphere \cite{Prahar_1} and $\lambda = \f{(d-2)(d-4)}{4}$. Now, the de Sitter momenta can be expressed in terms of flat space momenta as
\begin{align}
    &(k^\mu)_{\text{dS}} \to (k^\mu)_{\text{flat}} + \dfrac{\lambda}{2\delta^2} (\vec{k_i})_{\text{flat}}\\
    &(p_i^\mu)_{\text{dS}} \to (p^\mu_i)_{\text{flat}} + \dfrac{(d-2)\omega^2}{2\delta^2 E^2_i} (\vec{p_i})_{\text{flat}}.
\end{align}
We now expand the leading soft factor in $\omega$ and $\delta$ as
\begin{equation}
\label{sf}
    \lim_{\omega\to 0}\ \omega A\big(\{p_i\},\omega\hat{k} \big)= \sum_{i=1}^{n}e_i\ \dfrac{p_i^\mu\varepsilon_{\mu}}{ (p_i\cdot \hat{k})} \bigg[1-\dfrac{\lambda}{2\delta^2} \dfrac{\Vec{p}_i\cdot \bold{\hat{k}}}{p_i \cdot \hat{k}} \bigg].
\end{equation}
We can also write the sub-leading soft theorem including the perturbative corrections as
\begin{align}
\label{slf}
\lim_{\omega\to 0}\ \big(1+\omega\partial_\omega \big)A(\{p_i\},\omega \hat{k}) &= \sum_{i=1}^{n} e_i\bigg[- i\dfrac{\varepsilon_{\mu}\hat{k}_\nu J_i^{\mu\nu}}{p_i\cdot \hat{k}} + \dfrac{(d-4)}{4\delta^2} \dfrac{p_i^\mu \varepsilon_{\mu}}{(p_i\cdot \hat{k})^2} - \dfrac{(d-4)}{4\delta^2} \dfrac{p_i^\mu \varepsilon_{\mu}}{(p_i\cdot \hat{k})^2} p_i\cdot\partial_{p_i} \nonumber\\ &- \dfrac{\lambda}{2\delta^2} \dfrac{p_i^\mu \varepsilon_{\mu}}{p_i\cdot\hat{k}} \mathbf{\hat{k}}\cdot\partial_{\Vec{p}_i} 
- \dfrac{\lambda}{2\delta^2} \dfrac{(p_i^\mu\varepsilon_{\mu})(\vec{p}_i\cdot\mathbf{\hat{k}})}{(p_i\cdot\hat{k})^2} \hat{k}\cdot\partial_{p_i}
\bigg] + \dfrac{1}{\delta^2} \bar{M}^{(2)}_{n+1}(\{p_1,...,p_n\}, \hat{k}),
\end{align}
where $M^{(2)}_{n+1}=\bar{M}^{(2)}_{n+1} \Gamma_n$. To compute the perturbative corrections to Low's soft theorem, we need to determine $M^{(2)}_{n+1}$ using gauge invariance explicitly in the momentum space. The gauge invariance at this order becomes highly complicated and is therefore out of the scope of the present work. One can see from \eqref{sf} and \eqref{slf} that the corrections to the leading and sub-leading soft factors vanish in $d=4$ dimensions. This is consistent with the results of \cite{Sayali_Diksha}.

\section{Discussion}
\label{discussion_sec}
In this paper, we have studied the scattering amplitude of an arbitrary number of massless scalars and one soft photon. Similarly to flat space, we observed that the amplitude factorizes into two parts when the external photon becomes soft. We computed the $\frac{1}{\ell^2}$ correction to the leading and sub-leading soft photon theorems in this context. To setup our problem, we first gave a brief description of the scattering region inside the static patch of de Sitter spacetime. We then briefly discussed the modes of the  U(1) gauge field, which has been described at length in \cite{Sayali_Diksha}. The new thing in our setup is to describe the massless scalar mode and propagator in $d$-dimensions. We derived the orthonormal set of modes in the limit of large curvature and used it to derive the propagator. Then, we followed the same machinery as in \cite{Sayali_Diksha, SCB} to define the $\mathcal{S}$-matrix from the LSZ formalism. Given the $\mathcal{S}$-matrix, we computed the perturbative corrections to the leading and sub-leading soft factors. It is instructive to note that the massless limit of soft factors remain unchanged for different sets of scalar modes. Indeed, in \cite{Sayali_Diksha}, they used two different massive scalar modes and the mass-independent soft factors at the sub-leading order remain unchanged. In our work, we deduced a new set of massless modes and even then, the sub-leading soft factors still remain unchanged. It therefore points to the fact that these soft factors are expected to be universal.
\\~\\
Let us now discuss a couple of interesting future directions. First of all, one can try to understand the connection between the perturbative corrections of the soft factor and the asymptotic symmetries in the static patch of de Sitter spacetime. The non triviality in this direction is to first setup the $(d-2)$-dimensional sphere coordinates, which has been done in some setting in \cite{Prahar_1}. Once this is achieved conveniently in the de Sitter framework, one can try to derive the corrections to the Ward identities of large gauge transformations, as a result of the corrections of the soft factors that we have. One needs to be careful about the soft and hard charges in $d$-dimensions as the Ward identity requires a $d$-dimensional integral to be performed, see (3.11) of \cite{Prahar_1} and this integral in even and odd dimensions is slightly different. We have tried to analyze this in the present context but find it highly non trivial to arrive at the desired Ward identity for some arbitrary $(d-2)$-dimensional sphere function $\epsilon(x)$. This is because the integral is very complicated to do and we could not get a closed form analytic result for our corrected soft factors. We thus leave this task as a future direction. 
\\~\\
Another interesting direction could be to compute loop level corrections to the flat space soft factors in de Sitter spacetime. As is well known \cite{Laddha_Sen}, the loop level soft factors give logarithmic terms in the energy of the soft photon/graviton. How do these log terms get corrected is a question that deserves to be looked at. We would like to address this in the future. 

\section*{Acknowledgements}
PC thanks Divyesh N Solanki for suggesting corrections on a version of this draft and for various conversations about this work. PC is supported in part by the National Natural Science Foundation of China (Grant No. 12475105).

\end{document}